\documentclass[prb,twocolumn,amsmath,amssymb,floatfix]{revtex4}
\usepackage{graphicx}

\newcommand{\be}{\begin{eqnarray}}
\newcommand{\ee}{\end{eqnarray}}
\newcommand{\bfr}{{\bf r}}
\newcommand{\bfq}{{\bf q}}
\newcommand{\bfp}{{\bf p}}

\newcommand{\bfP}{{\bf P}}
\newcommand{\bfi}{{\bf i}}
\newcommand{\bfR}{{\bf R}}
\newcommand{\bfs}{{\bf s}}

\newcommand{\wbe}{\begin{widetext}}
\newcommand{\wee}{\end{widetext}}
\newcommand{\oncite}{\onlinecite}

\begin{document}
\draft

\title{Quantum fluids of self-assembled chains of polar molecules}

\author{Daw-Wei Wang$^{(1)}$, Mikhail D. Lukin$^{(2)}$,
and Eugene Demler$^{(2)}$}

\address{$^{(1)}$Physics Department, National Tsing-Hua University, Hsinchu,
Taiwan, ROC
\\
$^{(2)}$Physics Department, Harvard University, Cambridge, MA 02183
}

\date{\today}

\begin{abstract}
We study polar molecules in a stack of strongly confined pancake
traps.  When dipole moments point perpendicular to the planes of the
traps and are sufficiently strong, the system is stable against
collapse but attractive interaction between molecules in different
layers leads to the formation of dipolar chains,
analogously to the chaining phenomenon in classical rheological
electro- and magnetofluids. We analyze properties of the resulting
quantum liquid of dipolar chains and show that only the longest chains
undergo Bose-Einstein condensation with a strongly reduced
condensation temperature. We discuss several experimental methods for
studying chains of polar molecules.
\end{abstract}


\maketitle
Recent progress in trapping and cooling of chromium atoms
[\onlinecite{Cr}] and polar molecules [\onlinecite{cool_molecule}]
opened new directions for studying quantum systems with dipole
interactions. Manifestations of dipolar interactions have been
observed in the density profiles of the chromium
condensate\cite{dipole_size} and predicted in the
dispersion of the Bogoliubov mode [\oncite{dipole_simple}]. Exotic
many-body states arising from the anisotropy and long range
character of dipolar interactions have also been theoretically
proposed [\oncite{dipole_exotic,buchler}]. An important aspect of dipolar
systems is the attractive part of the interaction for dipoles
aligned head to tail, which can lead to the system collapse
[\oncite{collapse}]. 
To circumvent this problem, most theoretical
analysis focused on two and one dimensional systems with dipolar
moments perpendicular to the plane(axis) of the sample, in which
case the attractive part of the interaction is absent.

\begin{figure}
\includegraphics[width=7.5cm]{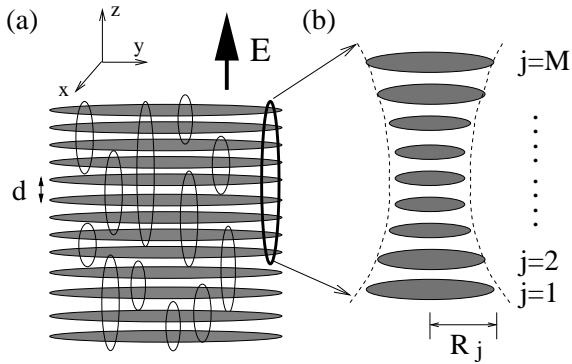}
\caption{
(a) Schematic figure of a stack of pancake traps. External electric field
polarizes molecules in the direction perpendicular to the layers. Vertical
ellipses denote self-assembled chains of molecules. (b) A typical shape of
a dipolar chain. Shaded ellipses denote single molecule wavefunctions
in individual layers (see Eq. (\ref{wf})) with radius $R_j$. 
}
\label{multilayer}
\end{figure}

In this paper we consider polar molecules in a three dimensional trap
which was divided into a stack of two-dimensional(2D) pancakes by
applying a strong one-dimensional optical lattice
[\oncite{2D_layer}](see Fig.  \ref{multilayer}(a)). When dipolar
moments are polarized in the direction perpendicular to the planes, 
the intra-layer interaction is
purely repulsive [\oncite{buchler}]. The attractive interaction, however,
is present for dipoles located on top of each other in different
layers. The system collapse in the direction of attractive
interaction is prevented by the strong optical
potential. In the remainder of this paper we focus on electric
dipolar interactions between polar molecules but our conclusions
should also apply to magnetic dipolar atoms [\oncite{Cr}].

Our main  observation is that
beyond a certain critical value of the dipole moments, attractive
interaction between molecules in different layers can bind them into
chains. Longer chains appear first and have a stronger binding
energy (see Fig. 2). At finite temperature there is a competition
between entropy that favors shorter chains and energy which is
minimized for longer chains. This leads to a {\it non-monotonic}
dependence of the distribution function on the length of the chains
(see Fig. \ref{N_a_T_fig}). This behavior of the
dipolar chains liquid (DCL) strongly resembles rheological electro-
and magnetofluids [\oncite{x}]. In both cases the dipolar
interaction leads to self-organization of elementary objects
(nanoscale grains in rheological fluids and molecules in our case)
into chains of varying length. In the rheological fluids
several finite temperature phase transitions have been predicted 
theoretically or numerically [\oncite{classical_dipole_chain}]. 
The chaining phenomena can also be directly 
imaged in a 2D thin film [\oncite{dipole_chain_2D}].
The new feature of DCL in polar molecule systems is the
possibility of Bose-Einstein condensation of chains at low
temperatures. We find that the longest chains condense first and
that in systems with many layers the condensation temperature is
strongly suppressed compared to the transition temperature of
individual molecules in one layer (insert to Fig.3). Such
suppression can be understood by observing that the presence of
chains of different lengths reduces the number of dipoles available
for the longest chains and hence limits the number of particles
available for the condensation. 
Finally we discuss implications of our results for experiments.

We start by investigating the structure of a single chain in a
multi-layered system in Fig. 1. For a strong 1D optical lattice there
is no spatial overlap of wavefunctions of molecules localized in
different layers, hence $M$ dipoles in successive layers can be
described by the Hamiltonian:
\be
H_M&=&\sum_{j=1}^M \left[\frac{\bfp_j^2}{2m}
+\frac{m\omega^2}{2}\bfr_j^2\right]
+\frac{1}{2}\sum_{i\neq j}^{M}V_{|i-j|}(|\bfr_{i}-\bfr_{j}|)
\nonumber
\ee
Here $V_j(\bfr)=D^2\left(|\bfr|^2-2j^2 d^2\right)/
\left(|\bfr|^2+j^2 d^2\right)^{5/2}$ is the dipole-dipole
interaction with $D$ being the electric dipolar moment in c.g.s.
units. $\omega$ is the in-plane trapping frequency and $m$ is the
mass of a single molecule. When $L$ is the total
number of layers in the 1D optical lattice, we can have
$M=2,3,\cdots,L$ as possible chain lengths. For simplicity we assume
that all layers are equivalent and contain the same number of
molecules. 
We define new variables: $\bfq_j\equiv \bfp_j-\bfP/M$ and
$\bfs_j\equiv\bfr_j-\bfR$, where $\bfP\equiv\sum_{j=1}^M\bfp_j$ and
$\bfR\equiv \frac{1}{M}\sum_{j=1}^M\bfr_j$ are the total momentum
and the center-of-mass(CM) position of the chain. 
As a result, $H_M$ can be separated into sum of the CM part 
($H_{\rm CM}=\frac{\bfP^2}{2Mm}+\frac{1}{2} Mm\omega^2 \bfR^2$) and
and the part that describes the
relative motion inside the chain:
\be
H_{\rm rel}&=&\sum_{j=1}^M \left[\frac{\bfq_j^2}{2m}+
\frac{m\omega^2\bfs_j^2}{2}\right]
+\frac{1}{2}\sum_{i\neq j}^{M}V_{|i-j|}(|\bfs_{i}-\bfs_{j}|).
\nonumber
\ee
The commutation relation between the new coordinates are:
$[\bfR,\bfP]=i\hbar$,
$[\bfs_j,\bfq_l]=i\hbar\delta_{jl}\left(1-M^{-1}\right)$, and all
other commutators are zero.
As we demonstrate below, important properties of a DCL are dominated
by the long chains ($M\gg 1$). So from now on we will treat ${\bf
s}_i$, ${\bf q}_j$ as canonically conjugate variables for chains of
all lengths. To calculate the binding energy from $H_{\rm rel}$, we
take the following variational wavefunction
\be
\psi_M(\{\bfs_{j}\})&=&\prod_{j=1}^M\frac{\exp(-|\bfs_{j}|^2/2R_j^2)}
{\sqrt{\pi}R_j},
\label{wf}
\ee
where the wavefunction width, $R_j$, is determined 
by the competition of attractive part of dipole interaction,
which favors localized wavefunctions, and the kinetic energy, which
opposes localization. Near the chain center interaction has
contributions from both sides of the chain, while
near the chain tips interaction comes from one side only and is
weaker. Therefore we expect a smaller size of the wavefunction near the
center compared to the tips of the chain (see Fig. \ref{multilayer}(b)).  
For simplicity we assume 
$R_j=R_0\left(1+\xi|j-(M+1)/2|^2\right)$, where $R_0$ and $\xi$
are two variational parameters to be obtained by minimizing
$\langle \psi_L(\{\bfs_{j}\}) |H_{\rm
rel}|\psi_L(\{\bfs_{j}\})\rangle$. 
Taking more sophisticated trial wavefunctions will not 
have a qualitative effect on the properties of DCL that
we discuss in this paper.

\begin{figure}
\includegraphics[width=8cm]{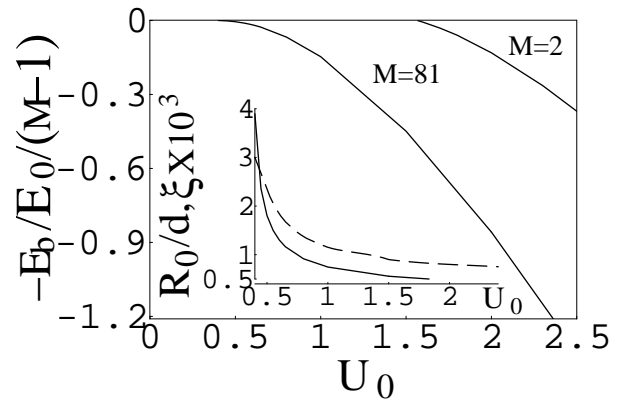}
\caption{Binding energy (note the minus sign) per molecule as
a function of dipole strength, $U_0$, for chain length $M=2$ and
$M=81$. Results are obtained using the variational wavefunction in
Eq. (\ref{wf}), with $\omega=0.01 E_0$. 
$E_0\equiv \hbar^2/md^2=\frac{2}{\pi^2}E_R$ with $E_R$ being the
recoil energy of the 1D optical lattice. Inset: Optimal values of
the variational parameters for a single chain, 
$R_0$(solid curve) and $\xi$ (dashed
curve), as a function of $U_0$ for $M=81$. Note that the value of
$\xi$ has been multiplied by $10^3$ to fit the scale of the vertical
axes. } 
\label{Eb_U0_fig}
\end{figure}
In Fig. \ref{Eb_U0_fig} we compare binding energies (per molecule)
of the shortest ($M=2$) and longest ($M=81$) chains in a system with
$L=81$ layers. The control parameter is the dimensionless strength
of the dipole interaction, $U_0\equiv m D^2/\hbar^2 d$, which can
be tuned either by increasing the electric field (which increases
the value of dipole moments of molecules), by changing the
interlayer distance $d$, or by increasing the effective
mass of molecules using an additional in-plane optical lattice. A
non-trivial result of our analysis is that bound states of many
molecules (long chains) appear first and have a stronger binding
energy. In the system under consideration, the longest chain
($M=L=81$), can exist when $U_0$ exceeds a critical value
$U_{0,L}^\ast\sim 0.4$. A bound state of two molecules appears only
for $U_0>U_{0,2}^\ast\sim 1.6$ [\oncite{resonance_note}]. So for
$U_0< U_{0,L}^\ast$ no chains are formed, for $U_0>U_{0,2}^\ast$ chains
of all lengths are stable, and for intermediate dipole strengths
($U_{0,L}^\ast<U_0<U_{0,2}^\ast$) there is a critical value of the
chain length $M_0$ such that chains of length $M_0$ or larger are
stable but shorter chains are not (except for the unbound dipoles,
$M=1$).  In the inset of Fig.
\ref{Eb_U0_fig}, we show the optimal values of the variational
parameters, $R_0$ and $\xi$, for the longest chain ($M=81$) obtained
as a function of $U_0>U_{0,L}^\ast$. Our results show that when
$U_0$ is small (weakly bound chains), the minimum width of the chain
can be several times larger than the interlayer distance, $d$. In
this limit the curvature of the shape, $\xi$, is large and the chain
has a shape of a horn in both ends 
(Fig. \ref{multilayer}(b)). For large dipole
strengths we find  tightly bound chains in which both $R_0$ and
$\xi$ are very small.

We now proceed to discuss the thermodynamics of DCLs. We start   
by making several simplifying assumptions. Specifically, we neglect 
repulsive interactions between chains and do not include bending 
of individual chains. The first assumption is justified rigorously 
in the limit of small number of molecules within each layer. For $N$ 
molecules in one layer, the characteristic strength
of the in-plane dipolar energy is $E_{\rm int} \sim N D^2/a^3_{\rm
osc}$, where $a_{\rm osc}=(\hbar/m\omega)^{1/2}$ is the in-plane
oscillator length. We can neglect $E_{\rm int}$ compared to $\hbar
\omega$ when $N^2\omega < (\hbar^5/D^4m^3)^{1/2}$.  
We further note that most conclusions involving thermodynamic properties 
of DCL will remain qualitatively similar even when interactions between
chains become important so long as they are not strong 
enough to lock molecules 
within the planes into a Wigner crystal phase [\oncite{buchler}].
Neglecting bending of individual chains can be
justified when temperatures are smaller than the characteristic
binding energy and the chain lengths are not too large. This
assumption may not necessarily apply near the condensation temperature
for modest dipolar moments (see Fig. 3 and discussion below)
but should be valid for large values of $U_0$ [\oncite{bending}].

When the interactions between chains are neglected, the
total energy of a DCL is a sum of energies of individual chains. 
Neglecting the bending modes, we can write the single chain energy
as a sum of the center of mass and binding energies:
$\varepsilon_{M,\bfi}=\hbar\omega(i_x+i_y+1)-E_{b,M}$. Here
$\bfi=(i_x,i_y)$ are quantum numbers for the CM coordinate in a 2D
harmonic trapping potential and $E_{b,M}$ is the binding energy of a
chain of length $M$. To understand thermodynamics of the DCL we need
to sum over all chain configurations with the constraint that the {\it
total} number of dipolar molecules is fixed. We enforce the constraint
using the Lagrange multiplier method and obtain he partition
function, ${\cal Z}$,
\be
\log {\cal Z}&=&-\sum_{M=1,M_0}^L\sum_{\bfi}
\sum_{\alpha_M}\log\left[1-
e^{-\beta(\varepsilon_{M,\bfi}-M\mu)}
\right],
\label{partition}
\ee
Here $\beta=1/k_BT$ and index
$\alpha_M=\frac{M+1}{2},\cdots,L-\frac{M-1}{2}$ counts possible CM positions of
chains in the $z$ direction.
\begin{figure}
\includegraphics[width=8cm]{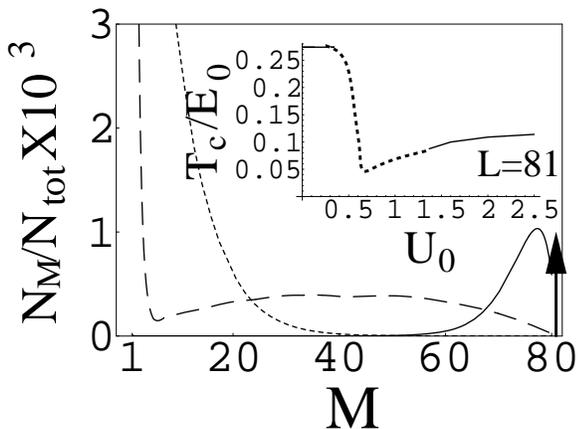}
\caption{Number of chains in a multilayer system ($L=81$) as a function of
the chain length, $M$. Other parameters are $N_{\rm tot}=10^5$, $U_0=2$, and
$\hbar\omega=0.01 E_0$. Condensate $T_c$ is $0.11 E_0$ for such parameters.
Solid, dashed, and dotted lines are for temperature
$T=0.1,$ 0.3, and 0.5 $E_0$ respectively, and the upward arrow indicates
the condensation of the longest chain ($M=L=81$)
Inset: Condensate $T_c$ as a function of dipole
strength, $U_0$, for multilayer system $L=2$ and $L=81$. 
The number of dipoles per layer for these two systems are set the same 
so that they have the same $T_c$ in noninteracting regime.
Solid lines are calculated within the noninteracting chain
approximation (discussed in the text). Dotted lines are schematic 
since reliable theory in the vicinity of $U_{0,L}^\ast$ should 
include interactions between chains.
}
\label{N_a_T_fig}
\end{figure}
In the above partition function the chemical potential
term $M\mu$ is proportional to the chain length $M$. This follows from
the fact that the numbers of chains are not fixed but can change in
dis/as-sociation type processes even in equilibrium. 
The only physical chemical potential
$\mu$ is that of individual molecules and controls the distribution of
chain lengths in thermal equilibrium.  This is similar to well known
systems of multi-ionized plasmas [\oncite{Landau_statistics}] and
classical rheological fluids [\oncite{x}]. To find $\mu$
in (\ref{partition}) we relate it to the  total
number of molecules $N_{\rm tot}=k_BT\partial\ln{\cal Z}/
\partial\mu=\sum_{M=1,M_0}^{L}MN_M$,
where $N_M$ is the number of chains of length $M$. At temperatures above
Bose-Einstein condensation we have 
\be
N_M(\mu)=(L-M+1)\left(\frac{k_BT}{\omega}\right)^2Li_2(z_M).
\label{N_M}
\ee
Here $z_M\equiv\exp\left[\beta(E_{b,M}+M\mu-\hbar\omega)\right]$ is
the fugacity of chains of length $M$, and
$Li_2(z)\equiv\sum_{k=1}^\infty\frac{z^k}{k^2}$ is the polylogarithm
function. In deriving expression (\ref{N_M}) we replaced summation
over discreet levels of the in-plane trap by the integration and used
the identity, $\int_0^\infty\int_0^\infty
\left[z^{-1}e^{x+y}-1\right]^{-1}dxdy=Li_2(z)$.  The prefactor,
$(L-M+1)$, counts the number of possible CM positions in the $z$
direction.  Bose-Einstein condensation means the appearance of a
macroscopic occupation of some microscopic state.  In our case this
happens when the argument inside the $\log$ of Eq.
(\ref{partition}) goes to zero for some particular value of the
quantum numbers $M$ and ${\bf i}$.  From the dependence of
the binding energy $E_{b,M}$ on the chain length $M$, one can see that
the condensation occurs for the longest chains, $M=L$, in the state
$i_x=i_y=0$.  To find the transition temperature we need to solve the
equation $\mu(T_c)= L^{-1}(\hbar\omega-E_{b,L}) \equiv \mu_c$.  For
temperatures below $T_c$ the chemical potential stays fixed at $\mu_c$
and the number of chains in the condensate is given by the equation
$N_c=L^{-1}\left[N_{\rm tot}-\sum_{M=1,M_0}^{L-1}MN_M(\mu_c)\right]$.

In Fig. \ref{N_a_T_fig} we show the calculated number of chains,
$N_M$, as a function of their length for different
temperatures. We choose  $U_0=2.0$ which is large enough to 
allow formation of the chains of all length (see Fig. \ref{Eb_U0_fig}).  
At high temperatures(dotted line), however, 
the entropy is more important, so short chains and
unbound molecules dominate. One can show that
in this limit the distribution function obeys the Saha's equation
that was originally derived for multi-ionized plasma
[\oncite{Landau_statistics}]
\be
N_M\sim \left(\frac{N_{\rm tot}}{L}\right)^M
\left(\frac{\hbar\omega}{k_BT}\right)
^{2M-2} e^{-E_{b,M}/k_BT}.
\ee
As the temperature is lowered to be of the same order as the
binding energy, the competition
between entropy and energy leads to a {\it nonmonotonic} distribution
of $N_M$ (dashed line), which is a special feature of the DCL approaching the
the regime of quantum degeneracy. 
When the temperature is reduced further (solid
line), quantum statistics becomes important and we find a strong
enhancement of the population of longer chains and eventually
Bose-Einstein condensation of the longest chains.

In the inset of Fig.
\ref{N_a_T_fig}, we show the condensate $T_c$ as a
function of the dipole strength for a multilayer ($L=81$) system. 
Solid lines in the $U_0\to 0$ regime are obtained treating molecules 
as noninteracting particles. For large $U_0$ regime we include 
self-assembly of molecules into chains but neglect interactions between chains.
In the intermediate regime of $U_0$ (around the
critical dipole strength, $U_{0,L}^\ast$),
we expect the approximation of noninteracting chains to break down
due to a 
larger transverse size of the chains (see the inset of Fig. \ref{Eb_U0_fig}).
We use dotted lines to schematically describe the expected results
of $T_c$ in this regime. We anticipate a sharp kink
in the transition temperature in analogy to
mixtures of bosonic atoms
near Feshbach resonance regime [\oncite{stoof}], 
due to the formation of real space bound states (chains) at $U_{0,L}^\ast$.
We point out that $T_c$ is strongly suppressed in a multilayer system
for $U_0\sim U_{0,L}^\ast$ because the number of the longest chains 
is very small. This is a special feature of dipolar chain liquid,
since in this regime binding energies of the longest chain
are small so that molecules are spread out among chains of various lengths.

We now discuss experimental issues related to the realization and
detection of the dipolar chain liquid proposed in this paper.
For typical polar molecules with electric dipole moment $D\sim 1$ Debye,
$m\sim 100$ atomic mass unit, and trapped in the optical
lattice of Nd:YAG laser ($d=\lambda/2=0.53$ $\mu$m), we find
the dimensionless dipole strength, $U_0\equiv mD^2/\hbar^2 d$, is
about $2.8$, and exceeds the critical dipole strength needed for
the formation of dipolar chains (see Fig. \ref{Eb_U0_fig}).
$U_0$ can be further controlled by an external DC electric field ($E_{DC}$). 
E.g. by starting from a BEC of the non-polarized molecules at $E_{DC} =0$ 
one can drive a system into a DCL state by adiabatically increasing 
electric field. The resulting dipolar chain liquid can be observed 
by several different methods. For example, 
the frequencies of collective modes that are inhomogeneous along the stack
will change dramatically for $U_0>U_{0,L}^\ast$, due to the large 
change in the stiffness of the dipolar cloud associated with  the emergence 
of long chains. Likewise, RF spectroscopy  
can be used to probe the energy gap resulting from the chain 
formation. Successive absorption peaks shall be observable in the excitation 
spectrum, yielding a signature of 
chaining effect in the dipolar cloud. Finally, a long-range density 
correlation between dipoles within the chains can be measured via noise 
correlation techniques after the cloud is released from the in-plane trap.

We now discuss some novel avenues opened by this work.  Firstly,
starting with a fixed number of layers and approaching thermodynamic
limit by increasing transverse system size,
we expect that there will be a phase transition between a simple superfluid
and a DCL at $U_0=U_{0,L}^\ast$.
Secondly, by increasing the dipole strength even further,  
a first order transition from superfluid of chains to a Wigner crystal
phase should occur. In the Wigner crystal phase chains should
form a triangular 
lattice with a fixed distance between them [\oncite{buchler}]. 
Finally, when considering bending and/or vibration of long 
chains at finite temperatures, we expect the DCL to be similar 
to a classical ferroliquid with nontrivial thermodynamic 
properties. We expect that each of these avenues correspond to an 
intriguing  direction for future theoretical and experimental research.

In summary, we have predicted the chaining effect of polar molecules
confined in a 1D optical lattice. The ground state and thermodynamic
properties of the system can be qualitatively changed by the formation
of chains. These intriguing phenomena should be observable
in the experiments with ultra-cold polar molecules.

We acknowledge useful discussion with D.S. Fisher, D. DeMille, 
D. Nelson, and Ite Yu. This work was supported by NSC (DWW),
NSF Career award (ML), NSF grants DMR-0132874 (ED), and CUA.


\begin{thebibliography}{99}
\bibitem{resonance_note}
We note that the critical dipole strength for two layer ($M=2$) can be
rigorously calculated. Our calculation shows that
$U_{0,2}^\ast\sim 0.71$, about two times smaller than the result obtained
by variational approach.
\bibitem{bending} We expect that including bending modes would favor
the longest chains even more by introducing physically
distinct states, leading to a further reduction of $T_c$.
\end{thebibliography}
\end{document}